\documentclass[preprint]{elsarticle}

\usepackage{lineno,hyperref}
\modulolinenumbers[5]
\hypersetup{
    colorlinks=true, % set 'true' for colored links (removes boxes around links)
}

\usepackage{amssymb}
\usepackage{amsmath}
\journal{N.I.M. A}

\begin{document}

\begin{frontmatter}

	\title{Toward precise neutrino measurements: An efficient energy response model for liquid scintillator detectors}
	%\title{An efficient energy response model for liquid scintillator detectors}


	\author[label1]{Logan Lebanowski\corref{cor1}\fnref{cor2}}
	\author[label1]{Linyan Wan}
	\author[label1]{Xiangpan Ji}
	\author[label1]{Zhe Wang}
	\author[label1]{Shaomin Chen}

	\address[label1]{Department of Engineering Physics, Tsinghua University, Beijing, P. R. China 100084}
		\cortext[cor1]{loganleb@hep.upenn.edu}
		\fntext[cor2]{Present address: Department of Physics and Astronomy, University of Pennsylvania, Philadelphia, PA 19104-6396, USA}

	\begin{abstract}
Liquid scintillator detectors are playing an increasingly important role in low-energy neutrino experiments. 
In this article, we describe a generic energy response model of liquid scintillator detectors that provides energy estimations of sub-percent accuracy. 
This model fits a minimal set of physically-motivated parameters that capture the essential characteristics of scintillator response and that can naturally account for changes in scintillator over time, helping to avoid associated biases or systematic uncertainties.  
The model employs a one-step calculation and look-up tables, yielding an immediate estimation of energy and an efficient framework for quantifying systematic uncertainties and correlations.  
	\end{abstract}

	\begin{keyword}
		neutrino detector \sep detector response \sep liquid scintillator \sep attenuation length
\PACS 29.40.Mc \sep 02.70.Ns
%29.40.Mc 	Scintillation detectors
%02.70.Ns 	Molecular dynamics and particle methods under Computational techniques; simulations
%02.70.Uu 	Applications of Monte Carlo methods under Computational techniques; simulations
%07.05.Tp 	Computer modeling and simulation under Computers in experimental physics


	\end{keyword}

\end{frontmatter}


\section{Introduction}

\subsection{Liquid scintillator detectors}

Liquid scintillator detectors are widely used in low-energy neutrino experiments because of their high photon yield relative to Cherenkov detectors.  
The current generation of MeV-scale neutrino detectors based on organic liquid scintillator include Borexino~\cite{Alimonti:2008gc}, KamLAND~\cite{Eguchi:2002dm}, Double Chooz~\cite{Ardellier:2006mn}, RENO~\cite{Ahn:2010vy}, and Daya Bay~\cite{Guo:2007ug}.
These highly successful experiments usually apply energy thresholds no higher than the inverse beta decay threshold of 1.8~MeV, and have physics aims that are sensitive to the measurement of energy.
A number of liquid scintillator experiments have been proposed or are under construction, and are expected to make high-precision measurements and discoveries on various topics in neutrino physics.  These experiments include SNO+~\cite{SNO+:2015}, JUNO~\cite{JUNO:2016}, RENO-50~\cite{R50:2015}, Jinping~\cite{JP:2016}, LENA~\cite{Wurm:2011zn}, THEIA~\cite{Alonso:2014fwf}, and a multitude of very short baseline reactor experiments~\cite{Vogel:2015wua}.  
In addition to organic liquid scintillator detectors, noble liquid detectors utilize scintillation and are currently under highly active research, development, and implementation~\cite{Chepel:2012sj}.  Collectively, these scintillation detectors are used to measure properties of neutrinos, search for dark matter, and more.  

\subsection{Energy-sensitive neutrino physics at the $\mathrm{MeV}$ \!scale}

MeV-scale neutrinos are produced in abundance by ``free'' sources such as the sun and nuclear reactors.  
They are also produced in the earth and in supernovae, and provide a unique probe of the physics of these sources in addition to valuable measurements of neutrino characteristics.

One unknown characteristic of neutrinos is their mass hierarchy; that is, whether the third neutrino mass eigenstate ($\nu_3$) is heavier or lighter than the first two ($\nu_1$ and $\nu_2$).
JUNO has demonstrated that by measuring reactor antineutrino oscillation over an approximately 50-km baseline, it has the potential to determine the mass hierarchy from a detailed analysis of the measured energy spectrum~\cite{JUNO:2016}.
The use of the fine structure of the energy spectrum imposes a rigorous requirement of a $2.6\%/\sqrt{E(\text{MeV})}$ energy resolution~\cite{JUNO:2013}.

MeV-scale neutrinos also shed light on questions concerning the solar model.
The carbon-nitrogen-oxygen (CNO) fusion process generates neutrinos while fueling the sun.
A CNO neutrino measurement could provide a direct test of the solar-core metallicity~\cite{Haxton:2008}.
Jinping, a forthcoming neutrino observatory with the lowest cosmogenic and reactor antineutrino background, has the capacity to discover CNO neutrinos at a five sigma statistical significance, assuming a 500~p.e./MeV, or $4.5\%/\sqrt{E(\text{MeV})}$, energy resolution~\cite{JP:2016}.

These measurements will greatly depend on precise measurements of energy, for which a thorough understanding of detector energy response is necessary.

\subsection{Detector energy response}

The energy response of a detector refers to the typically nonlinear and spatially nonuniform dependence of the detector output on interaction energy and position within the detector.  The response can be modeled analytically or by Monte-Carlo simulation, with the preference depending on the accuracy and time requirements of the implementation.  
In either case, various data are needed to calibrate each model component.  It is also noteworthy that the response of scintillator, and possibly other detector components, changes over time.  
The current generation of liquid scintillator experiments model the energy response of their detectors as in Refs.~\cite{Bellini:2013lnn, Yoshida:2010zzd, Abe:2014bwa, Seo:2016uom, An:2016ses}.

This paper describes an energy response model of liquid scintillator detectors that employs both analytical and simulated components, and can be fit to data~\cite{Logan:2016}.  The simulation-based look-up tables can be computed quickly, and along with a one-step calculation, the model provides immediate energy estimates.  
Section~\ref{sec:model} describes the components of the energy response model.
Section~\ref{sec:validate} presents validation using simulations of two different detector geometries, and Section~\ref{sec:systematics} discusses systematics due to variations with time.  
Section~\ref{sec:conclusion} concludes the paper.

\section{Energy response model}
\label{sec:model}

\begin{figure}[htbp] %  figure placement: here, top, bottom, or page
	\centering
					% left bottom right top
		\includegraphics[trim=185 140 85 130,clip,width=0.64\columnwidth]{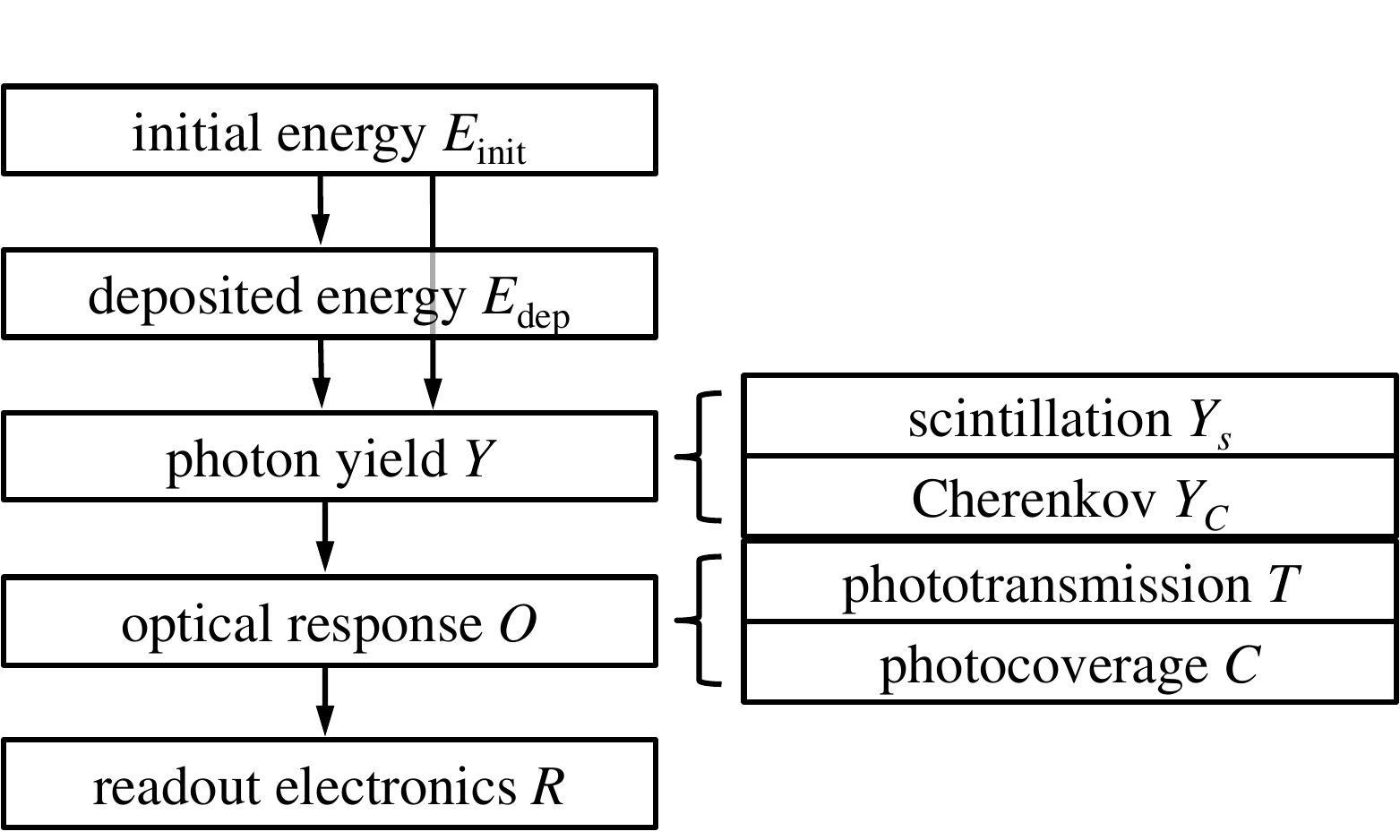}
	\caption{Flowchart of the detector response model.}
	\label{fig:flowchart}
\end{figure}

A general expression for the observed energy $E_\mathrm{obs}$ explicitly depends on the initial energy $E_\mathrm{init}$ and deposited energy $E_\mathrm{dep}$ of a particle in the scintillating volume, and the position $\vec x$ around where the particle induced scintillation: 
\begin{equation}\label{eq:nest}
	E_\mathrm{obs}(\vec x, E_\mathrm{init}, E_\mathrm{dep})=E_0\sum_{i=1}^NR_i\left\{O_i\left[\vec x, Y\left(E_\mathrm{init}, E_\mathrm{dep}\right)\right]\right\},
\end{equation}
where $Y$ represents the photon yield, $O$ represents the optical response, $R$ represents the single channel response of the readout electronics (RE), $N$ is the number of photosensor channels, and $E_0$ is a calibration constant that converts the sum to units of energy.  
Simply stated, this expression relates the observed energy to the distribution of signals induced on the array of photosensors.  It does not consider the \emph{distribution} of Cherenkov photons that \emph{directly} generate a signal in a photosensor, however, this is typically a small effect (see Section~\ref{sec:identicalness}) that could be incorporated with additional effort.  
We also note that the expression assumes a localized energy deposition, however, the response to energy deposited over a distance or at multiple vertexes, can be estimated with Eq.~(\ref{eq:nest}) applied at an appropriate number of positions $\vec x$ associated with the energy deposition.  
The basic flow of the detector response model is illustrated in Fig.~\ref{fig:flowchart}.

To obtain a simpler and more computationally efficient model, Eq.~(\ref{eq:nest}) may be modified so that it is not a sum over three nested functions but instead a product of three functions.  
For scintillation detectors, which yield approximately $10^{4}$ photons per MeV of deposited energy, the convolution of all the physical processes of detector response generally yields a Gaussian distribution, making this modification statistically feasible via the law of large numbers. 
The simplification can be achieved with two approximations: 
treating Cherenkov photons with the same wavelength-dependent transmission as scintillation photons and ignoring any position dependence of the RE response. 
The former approximation can be avoided by applying a distinct optical response for each type of photon, however the Cherenkov contribution is typically small.  
The latter approximation may be avoided by correcting each channel for $R_i$, in each event.  
These approximations lead to an implicit sum over the RE channels for $R$ and $O$, and are further discussed in the next section.  
With the approximations, the model is simplified to
\begin{equation}\label{eq:model}
	E_\mathrm{obs}(\vec x, E_\mathrm{init}, E_\mathrm{dep})=E_0\cdot R\cdot C(\vec x)T(\vec x) \cdot \left[Y_s+f\ Y_C\right](E_\mathrm{init}, E_\mathrm{dep}),
\end{equation}
where $Y$ is expressed with the two components $Y_s$ and $Y_C$ (scintillation and Cherenkov photon yields), the optical response function $O$ has been separated into two basic components $C$ and $T$ (photocoverage and phototransmission), and $R$ is the full-detector electronics response as a function of total incident charge (assuming no single-channel corrections).
The components introduced here are present in Fig.~\ref{fig:flowchart}.  

Though the two approximations have separated the components of the model so that they are no longer nested, the output of one component still depends on the input of another.  For example, if applying the model to data ($E_\mathrm{obs}$), the RE response should be corrected first, then optical response, and finally, scintillator photon yield.

\subsection{Photon yield}
\label{sec:photonYield}

The photons produced in a scintillator are from two main processes: scintillation and Cherenkov radiation.  The yields from both processes depend on the particle type and energy, and the properties of the scintillator.  For MeV-scale electrons in an organic liquid scintillator, Cherenkov photons typically contribute several percent to the total photon yield.  A neutral particle, such as a $\gamma$, can be treated by modeling its photon yield as a sum of over its charged daughters.  

\subsubsection{Scintillation yield}

The production of scintillation photons in an organic scintillator is often quantified with Birks' law~\cite{Birks:1951}, which is an empirical formula for the photon yield per distance traveled by the incident particle:
\begin{equation}\label{eq:Birks}
	\frac{dY_s}{dx}=A\frac{dE/dx}{1+k_B\cdot dE/dx},
\end{equation}
where $A$ is a normalization constant, $k_B$ is the Birks' constant of the scintillator, and $dE/dx$ is the energy loss of the incident particle, which depends on its energy.
Scintillation photons are emitted isotropically.

\subsubsection{Cherenkov yield}

In addition to scintillation light, charged particles with a mass $m$ generate Cherenkov radiation above an energy threshold of $mc^2/\sqrt{1-n^{-2}}$, where $n$ is the refractive index of the scintillator.
For electrons in Linear Alkyl Benzene (LAB), this threshold corresponds to about 0.17 MeV. 
The photon yield per distance traveled per unit wavelength $\lambda$ is~\cite{PDG:2014}
\begin{equation}\label{eq:Cherenkov}
	\frac{d^2Y_C}{dxd\lambda}=\frac{2\pi\alpha q^2}{\lambda^2}\left(1-\frac{1}{\beta^2n^2(\lambda)}\right),
\end{equation}
where $\alpha$ is the fine structure constant, $q$ is the charge of the particle in units of electron charge, and $\beta$ is the relative velocity of the particle, which depends on its energy.  Cherenkov photons are produced in a cone around the trajectory of the charged particle.  In scintillator, Cherenkov photons are often absorbed and can then be re-emitted isotropically.  

\subsubsection{Approximation of identicalness}\label{sec:identicalness}

Photons generated by the above two modes of radiation have different wavelength ($\lambda$) distributions.
As a result, their attenuation through the scintillator and photosensor acceptance are different. 
A distinct optical response can be applied for each type of photon, however since the Cherenkov contribution is typically small, scintillation and Cherenkov photons are assumed to experience identical optical processes. 
We denote the relative detection probability of Cherenkov to scintillation photons as $f(\lambda)$, and the average of $f(\lambda)$ over the transmission spectrum of the scintillator and the acceptance spectrum of the photosensors, as $f$. 
The resulting expression of the total photon yield is
\begin{equation}
	Y=Y_s+fY_C.
\end{equation}
Since the actual yield of scintillation photons [coefficient $A$ in Eq.~(\ref{eq:Birks})] may not be known, $f$ is also made to include the relative yield of Cherenkov to scintillation photons.  As such, the spectra of $Y_s$ and $Y_C$ are both normalized to 1.  
With this normalization, the relatively low Cherenkov photon yield and lower transmission typically result in $f = \mathcal{O}$(1\%) for MeV-scale electrons, likely yielding negligible uncertainty from the approximation of identicalness except within a few absorption lengths of the edge of the scintillator where the Cherenkov photons are unlikely to be absorbed.  The approximation also neglects photosensor sensitivity to wavelengths outside the range of scintillation.  

\subsubsection{Partial energy deposition}

The photon yield function $Y(E_\mathrm{init}, E_\mathrm{dep})$ explicitly depends on both initial and deposited energies to account for cases when a particle may not begin or end its energy deposition in the scintillator, such as near the boundaries of the scintillator volume. % or when originating from a deployed calibration source.
It is assumed that an experiment determines $Y$ for the case that $E_\mathrm{init} - E_\mathrm{dep} =$ 0, for example, by studying interactions at the center of the detector.  

To account for the case where $E_\mathrm{init}-E_\mathrm{dep}>0$, the following treatment can be applied.  As both $Y_s$ and $Y_C$ are determined by integrals over the particle's energy [see Eqs.~(\ref{eq:Birks}) and (\ref{eq:Cherenkov})], the dependence of $Y$ on the initial and deposited energies can be expressed as
\begin{equation}
	Y=\int_{E_\mathrm{init}-E_\mathrm{dep}}^{E_\mathrm{init}}\frac{dY}{dE}dE=\int_0^{E_\mathrm{init}}\frac{dY}{dE}dE-\int_0^{E_\mathrm{init}-E_\mathrm{dep}}\frac{dY}{dE}dE.
\end{equation}
Thus, the dependence of the photon yield on energy as determined with $E_\mathrm{init} - E_\mathrm{dep} =$ 0 can be utilized twice to account for the case when $E_\mathrm{init}-E_\mathrm{dep}>0$.

This treatment may be unnecessary when selecting events whose position reconstructs within a fiducial volume, in which case the first interaction of a particle would occur in the scintillator and it would deposit all of its energy in the scintillator; i.e., $E_\mathrm{init}-E_\mathrm{dep}=0$.  On the other hand, it may enable the analysis of events outside the fiducial volume or within a deployed source container.  In such analyses, a general solution to minimize any bias of the fitted central value is to use the `calorimeter function' described in Ref.~\cite{Cheng:2016ykr}.  

\subsection{Optical response}

The optical response $O_i(\vec x)$ is the probability that a photon generated at $\vec x$ propagates through the scintillator, makes contact with photosensor $i$, and generates a signal in it.
This probability is separated into two major components: phototransmission $T(\vec x)$, which is governed by an effective attenuation length, and photocoverage $C(\vec x)$, which depends on the photosensor arrangement and photosensitive area.  Both components are represented with look-up tables; i.e., maps.  

\subsubsection{Phototransmission}
\label{sec:transmission}

The fraction of photons that could reach a photosensor is determined by the properties of the propagation medium and the distance from the initial vertex to the photosensor.  
The probability that a photon of wavelength $\lambda$ will travel a distance $l$ through a medium of attenuation length $L(\lambda)$ follows Beer's law:
\begin{equation}
	T(\lambda, l)=\exp(-l/L(\lambda)).
\end{equation}
The fraction of $n$ emitted photons that is transmitted from production
vertex $\vec x$ to the continuous, physical surface that contains the photosensitive surfaces of the photosensors is then 
\begin{equation}\label{eq:transFraction}
	T(\vec x)=\frac1n\sum_{i=1}^n\exp\left(-\frac{l_i(\vec x)}{L}\right),
\end{equation}
where $L$ is an effective attenuation length that averages over the emission and attenuation spectra of the scintillator, and $l_i(\vec x)$ is the distance traveled by scintillation photon $i$ from the production vertex $\vec x$ to the continuous surface.
To separate the physics of the medium from the geometry of the detector, the attenuation length $L$ is separated from the photon distances $l_i(\vec x)$ by expanding each exponential and grouping terms of the same order:
\begin{equation}\label{eq:transmission}
	\begin{aligned}
		T(\vec x)=1+\sum_{k=1}^\infty\left(-\frac1L\right)^km_k(\vec x),\\
		m_k(\vec x)=\frac1n\cdot\frac1{k!}\sum_{i=1}^n\left[l_i(\vec x)\right]^k.
	\end{aligned}
\end{equation}
The sum $m_k(\vec x)$ is a map determined with simulation as outlined below.
The number of these moment maps needed for the calculation of $T(\vec x)$ is determined primarily by the size of the detector: if $l_i\ll L$ for nearly all $i$ at all $\vec x$, the first moment map ($k=1$) is sufficient.  The separation of $L$ from the $l_i(\vec x)$ simplifies the application of the model in that $L$ can be fit once to data and the moment maps generated once using simulation.  

A pseudocode example of how to generate transmission moment maps using a complete simulation of a detector is given below. 
In the example, photons are essentially used as geometric ray tracers, however wavelength-dependent reflection and refraction involving detector materials are included.  Scintillator scattering and absorption lengths should be made effectively infinite in the simulation.  
\begin{enumerate}
	\item Generate particles uniformly in the detector.
	\item For each particle, loop over all photons produced. 
	\item For each scintillation photon, loop over its vertexes, saving the position of its initial vertex and summing over the distance between consecutive vertexes in the scintillator per Eq.~(\ref{eq:transmission}).
\begin{enumerate}
	\item If the photon reached the continuous physical surface that contains the photosensors, save the summed photon distances through the scintillator (the distances are summed until the photon reaches the boundary for the final time). 
\end{enumerate}
	\item The transmission moment maps are the sums over the scintillation photon distances through scintillator, as a function of position. 
\end{enumerate}
The moment maps are primarily geometric and would need updating only if the detector geometry or a material reflectivity or transmissivity changed significantly.  With these moment maps, $L$ can be directly fit to data in the context of the complete model in Eq.~(\ref{eq:model}).  Demonstrations are given in Section~\ref{sec:validate}.  As a fit parameter, $L$ can be determined versus time, naturally accounting for changes in scintillator attenuation versus time.  

Detectors with more than one volume of scintillator can have a transmission map with a distinct attenuation length for each volume.
Following the same derivation as for Eq.~(\ref{eq:transmission}), the transmission map of two distinct scintillators is expressed as
\begin{equation}
	\begin{aligned}
		T(\vec x)=&1+\sum_{k=1}^\infty\left[\left(-\frac1{L_1}\right)^km_{1,k}(\vec x)+\left(-\frac1{L_2}\right)^km_{2,k}(\vec x)\right]\\
		&+\sum_{a,b}^{\infty}\left(-\frac1{L_1}\right)^a\left(-\frac1{L_2}\right)^bm_{ab}(\vec x),\\
		m_{j,k}(\vec x)=&\frac1n\frac1{k!}\sum_{i=1}^n\left[l_{j,i}(\vec x)\right]^k,\\
		m_{ab}(\vec x)=&\frac1n\frac1{a!}\frac1{b!}\sum_{i=1}^n\left[l_{1,i}(\vec x)\right]^a\left[l_{2,i}(\vec x)\right]^b,
	\end{aligned}\label{eq:2vtrans}
\end{equation}
where $a$, $b \ge 1$, and $l_{j,i}$ is the total length of the $i$th path through the $j$th medium such that $l_{1,i}+l_{2,i}=l_i$.
The $k$th-order transmission map consists of $m_{1,n}$, $m_{2,n}$, and all $m_{ab}$ for which $a+b=n$.
For this fit, it may be appropriate to allow two distinct photon yields, or energy scales, for the two volumes.
Equation~(\ref{eq:2vtrans}) may also be used to fit two distinct attenuation length components of a single scintillator; namely, the scattering and absorption lengths.
A value could be fit for each component by setting $l_{1,i}=l_{2,i}=l_i$.  Alternatively, if a value is known for one component, it could be fixed while the other is fit.

\subsubsection{Photocoverage}
\label{sec:coverage}

Of the total number of photons that could reach a photosensor, the fraction that does is largely determined by geometry; namely, the photosensor arrangement and area.  
Additionally, the fraction depends on the reflectivity or refractive indexes of the detector components.
Furthermore, the efficiency of the photosensors may exhibit dependencies on the incident angle of the photons, on the interaction position of the photons on the photosensor, and on the orientation of the photosensor relative to the local magnetic field.
Depending on the level of detail of the simulation, a generated photocoverage map can include all of these effects.
A pseudocode example of how to generate a photocoverage map using a complete simulation of a detector is given below. 
In the example, scintillator scattering and absorption are set to nominal to account for (wavelength-dependent) interactions with detector materials.
\begin{enumerate}
	\item Generate particles uniformly in the detector.
	\item For each particle, loop over all photons produced. 
	\item For each photon, loop over its vertexes, saving the position of its initial vertex.
\begin{enumerate}
	\item If the photon reached the continuous physical surface that contains the photosensors, count it as a `BOUNDARY' photon. 
	\item If the photon vertex generated a signal in a photosensor, count it as a `SENSOR' photon (these photons are also counted as `BOUNDARY' photons).  
\end{enumerate}
	\item The photocoverage map is SENSOR($\vec x$)/BOUNDARY($\vec x$). 
\end{enumerate}
The photocoverage map would need updating if a significant fraction of photosensors became inactive.  For less significant fractions, the model can be corrected to first-order by simply scaling to the fraction of active channels (e.g., see Ref.~\cite{An:2016ses}).  
Illustrations of a photocoverage map are shown in Section~\ref{sec:validate}.

\subsection{Readout electronics response}

The RE response, or $R$ in Eq.~(\ref{eq:nest}), is unique to each experiment.  For example, the RE may provide a single-valued measurement of the photosensor signal in units of analog-to-digital converter (ADC) channels, or provide an entire digitized signal waveform.  
In any case, this study assumes that each RE channel is calibrated and the response of the RE is linear.  
Furthermore, the signal selection time window of the electronics and the noise rates of the photosensors are neglected and therefore have no impact on the calibration of the model demonstrated here.

Finally, it is noted that any nonlinear description of RE channel response $R_i$ would ideally be used to `correct' the data read out from each RE channel $i$.  This would avoid one of the two approximations made to obtain the model as expressed by Eq.~(\ref{eq:model}), which was to ignore any position dependence of $R$.

\subsection{Energy scale}
The output of the RE need only be multiplied by energy scale constant $E_0$ to recover units of energy.  
For example, if $R$ has units of ADC/photo-electron, $E_0$ has units of MeV/ADC.  
Essentially, $E_0$ is the calibration of all detector components: the scintillator volume, photosensors, RE, {\it etc}.  
Thus, the value of $E_0$ depends on the energy of the calibration source, and on the location or distribution of the source.  
If fitting the energy spectrum of a source near the scintillator boundary or from within a deployed container, care should be taken to minimize any bias of the fitted central value~\cite{Cheng:2016ykr}.  
Finally, it is noted that the value of $E_0$ is also particle-dependent.

\section{Validation with simulation}
\label{sec:validate}

\subsection{Simulation setup}
\label{sec:simSetup}

This section validates the detector response model described in Section~\ref{sec:model} using a simulation based on the design of the Jinping neutrino experiment~\cite{JP:2016}.  
All simulations are performed with \textsc{Geant4}~\cite{Geant4:2016}.

\begin{figure}[htbp] %  figure placement: here, top, bottom, or page
	\centering			% left bottom right top
		\includegraphics[trim=30 0 210 0,clip,width=0.49\columnwidth]{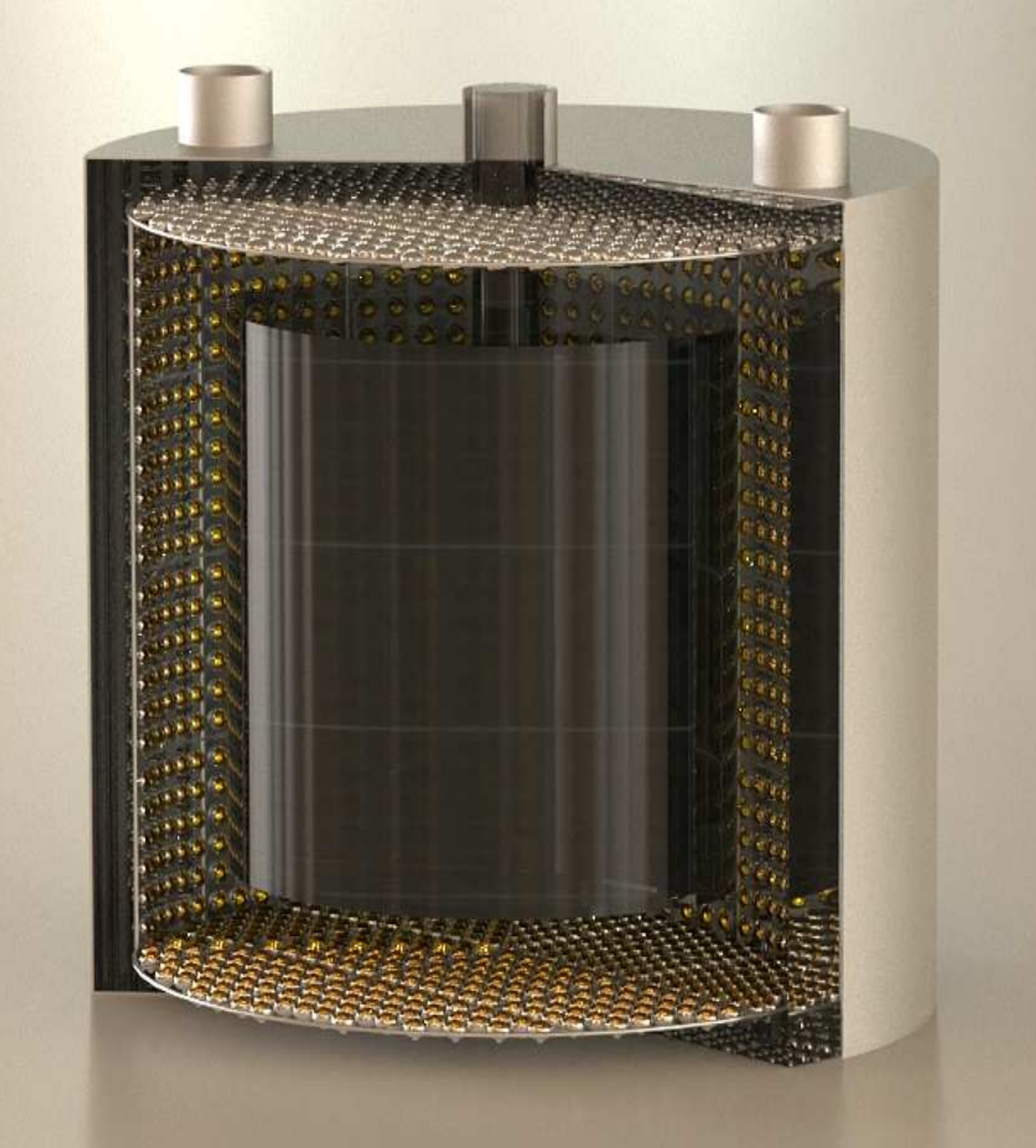}
		\includegraphics[trim=30 0 210 0,clip,width=0.49\columnwidth]{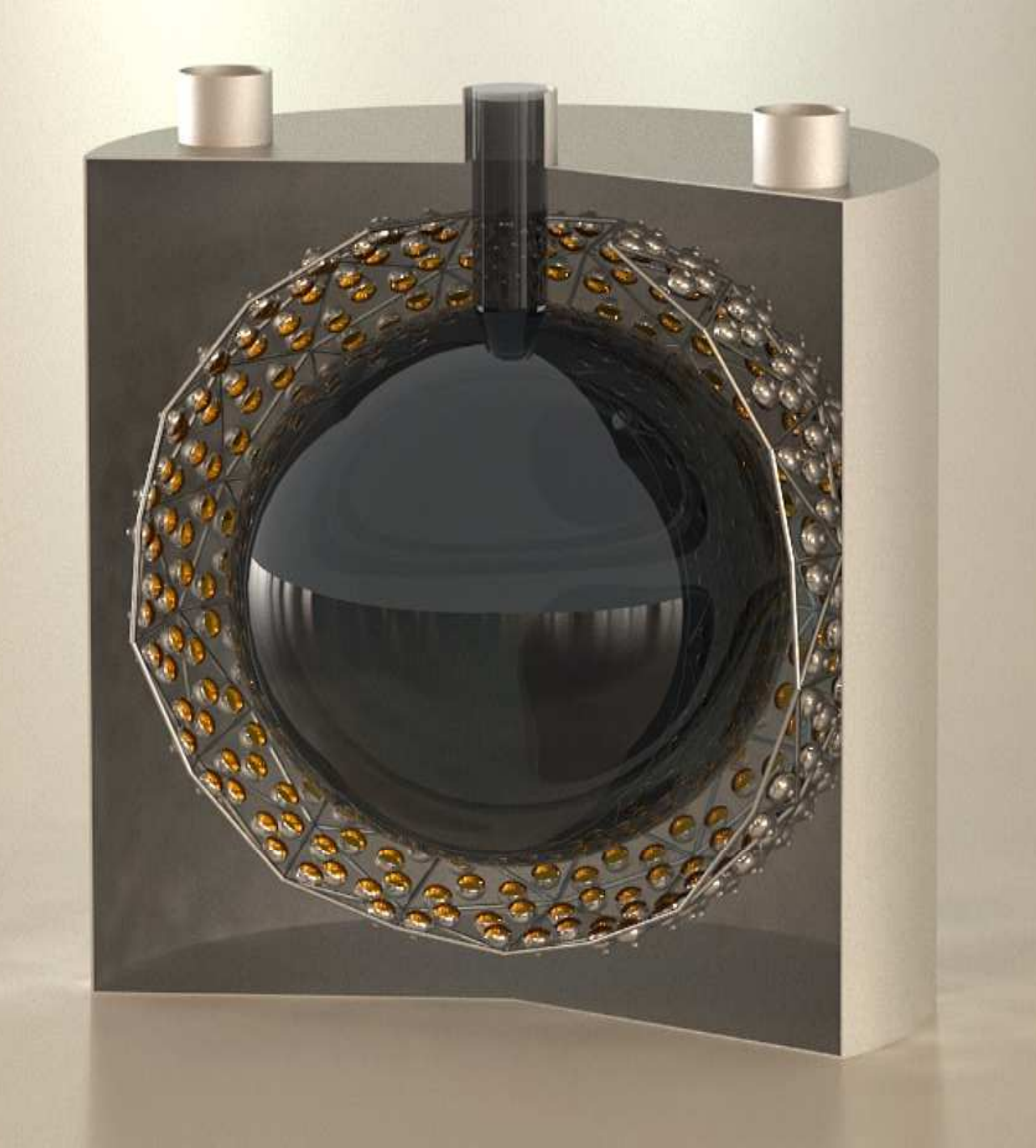}
	\caption{Simulation setup of a liquid scintillator detector in the shape of a cylinder (left) or sphere (right), based on those used in Ref.~\cite{JP:2016}.}
	 \label{fig:sim}
\end{figure}

To test the model with different detector shapes, both a spherical and a cylindrical kiloton liquid scintillator detector were simulated, as illustrated in Fig.~\ref{fig:sim}.
The radii and the half-height of the scintillator volumes are all 5.65~m.
Photomultiplier tubes (PMTs) are uniformly placed at 8.3~m from the detector center or central axis.
They are immersed in water, which is separated from the liquid scintillator by an acrylic vessel.
The outer cylindrical tanks are stainless steel, both with a radius and a half-height of 9.0~m.

The liquid scintillator is pure LAB, the properties of which have been reported in Ref.~\cite{LAB:2015}.  
In the simulations, the effective attenuation length of the LAB is 20.0~m, Birk's constant $k_B$ is effectively $0.065$~mm/MeV, and the photo-electron (p.e.) yields are 500 and 520~p.e./MeV for 16-MeV electrons at the centers of the cylindrical and spherical setups, respectively.
The PMT models are based on the Hamamatsu R3600-02 20-inch PMT~\cite{HamamatsuR3600}.  
There are 4085 and 2844 PMTs in the cylindrical and spherical setup, respectively.  The PMTs are assumed to be perfectly efficient.  
We take $R = 1$ in Eq.~(\ref{eq:model}) without applying units; thus, $E_0$ has units of MeV/p.e.

\subsection{Photon yield - nonlinearity}
\label{sec:photonYieldFit}

As described in Section~\ref{sec:photonYield}, scintillator photon yield is modeled by Birk's law and the well-known expression for Cherenkov radiation.
The energy scale $E_0$, Birk's constant $k_B$, and relative Cherenkov contribution $f$ can be simultaneously determined by fitting the observed energy $E_\mathrm{obs}$ of various energy sources at a single position ${\vec x}_\mathrm{cal}$.  In this case, the phototransmission and photocoverage maps $T({\vec x}_\mathrm{cal})$ and $C({\vec x}_\mathrm{cal})$ are constants, and therefore grouped with $E_0$ into a composite energy scale $E_{0,CT}({\vec x}_\mathrm{cal}) \equiv E_0\cdot C({\vec x}_\mathrm{cal})\cdot T({\vec x}_\mathrm{cal},L)$, which has units of MeV/photon.  Thus, Eq.~(\ref{eq:model}) reduces to 
\begin{equation}\label{eq:NL}
	E_\mathrm{obs}({\vec x}_\mathrm{cal},E_\mathrm{init})=E_{0,CT}({\vec x}_\mathrm{cal})\cdot [Y_s+fY_C](E_\mathrm{init}),
\end{equation}
where the deposited energy $E_\mathrm{dep}$ does not appear because it is assumed to always equal $E_\mathrm{init}$ at the chosen position, ${\vec x}_\mathrm{cal}$.   In fitting, both $Y_s$ and $Y_C$ are normalized to 1 (at $E_\mathrm{init} =$ 16~MeV) as mentioned in Section~\ref{sec:identicalness}.  This normalization modifies Eq.~(\ref{eq:NL}): 
\begin{equation}\label{eq:NLnorm}
	E_\mathrm{obs}({\vec x}_\mathrm{cal},E_\mathrm{init})=\check{E}_{0,CT}({\vec x}_\mathrm{cal})\cdot \left[\frac{\hat{Y}_s+f\hat{Y}_C}{1+f}\right](E_\mathrm{init}),
\end{equation}
where the hat ($\hat{Y}$) indicates normalization and the check ($\check{E}$) indicates that $E_{0,CT}$ absorbs the inverse of the scale to which $Y$ is normalized. This gives $\check{E}_{0,CT}$ units of MeV.  
Although $\check{E}_{0,CT}$, $k_B$, and $f$ can be fit simultaneously, it may be advantageous to determine $\check{E}_{0,CT}$ first, with a high-statistics measurement of a single reference source, and then fit for $k_B$ and $f$ using Eq.~(\ref{eq:NLnorm}) with $\check{E}_{0,CT}$ fixed.  This can also help reduce the impact of correlations in the fit.  

\begin{figure}[htbp] %  figure placement: here, top, bottom, or page
	\centering
		\includegraphics[width=0.49\columnwidth]{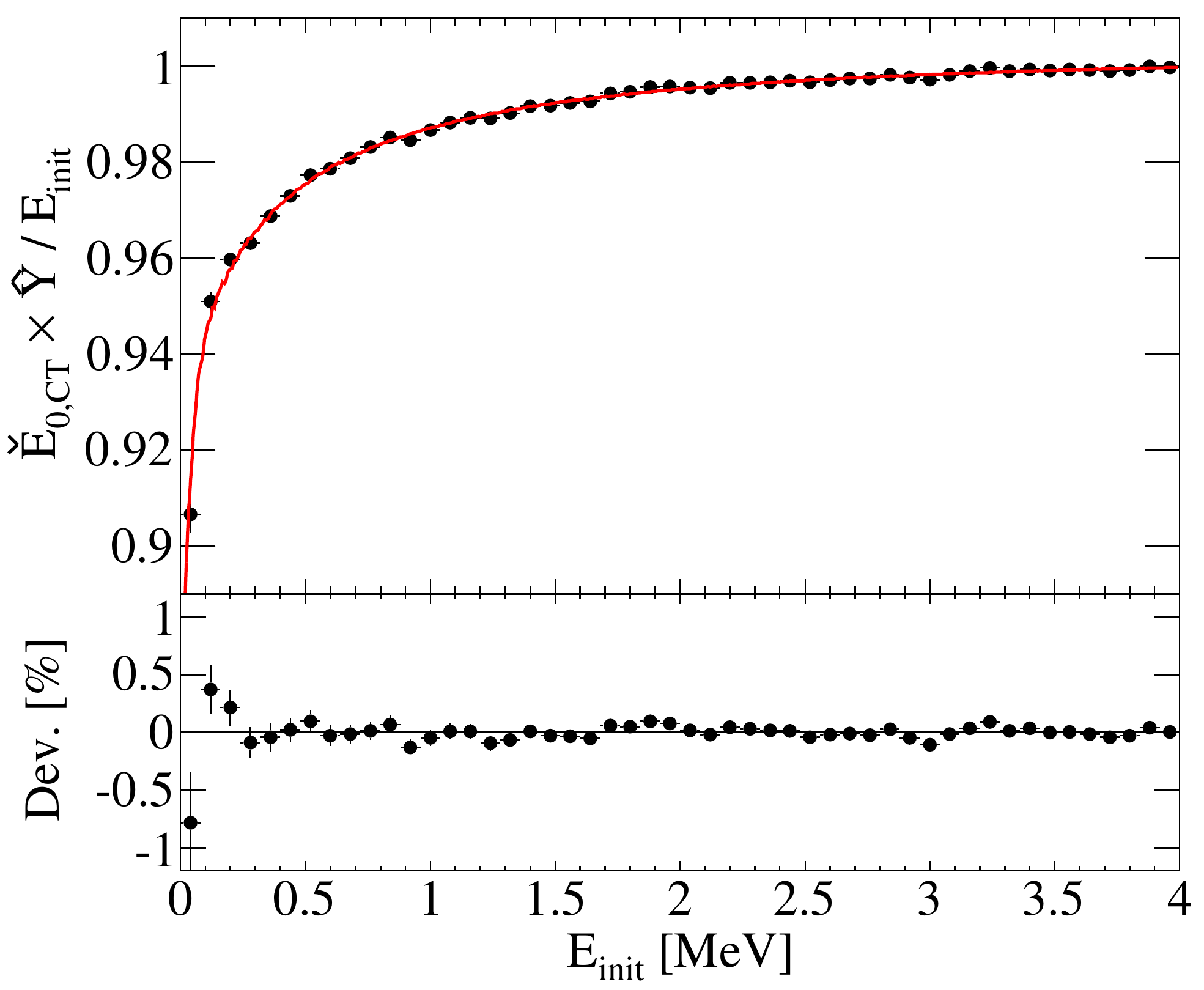}
	\caption{Simulated nonlinear response of liquid scintillator to electrons, fitted with Birk's law and a Cherenkov component.  The bottom panel shows the percentage error between the fitted curve and the simulation.}
	 \label{fig:quench}
\end{figure}

Figure~\ref{fig:quench} illustrates the nonlinear response of the scintillator for electrons in the center of the spherical detector and a fit of Eq.~(\ref{eq:NLnorm}) with $\check{E}_{0,CT}$ fixed to 16~MeV. 
The fit results are $k_B = 0.0657\pm0.0014$~mm/MeV and $f = 0.0246\pm0.0004$.  The accuracy of the model is illustrated in the bottom panel of Fig.~\ref{fig:quench}, which shows the percentage error between the simulated data and the fit to the model.  The root mean square (RMS) of the bottom panel is 0.14\%.  

Besides fitting for $\check{E}_{0,CT}$ individually, another approach to handling the correlations is a combined fit of the various energy sources at an additional position, yielding two distinct $\check{E}_{0,CT}$ and a single $f$ and single $k_B$.  
In the case of measurements at the center of the spherical detector ($r =$ 0) and at a radius of $r =$ 4.5~m, the combined fit function would be
\begin{equation}\label{eq:NL2}
	\begin{aligned}
		E_\mathrm{obs}(r=0,E_\mathrm{init})=\check{E}_{0,CT}(r=0)\cdot \left[\frac{\hat{Y}_s+f\hat{Y}_C}{1+f}\right](E_\mathrm{init}),
		\\	E_\mathrm{obs}(r=4.5~\mathrm{m},E_\mathrm{init})=\check{E}_{0,CT}(r=4.5~\mathrm{m})\cdot \left[\frac{\hat{Y}_s+f\hat{Y}_C}{1+f}\right](E_\mathrm{init}).
	\end{aligned}
\end{equation}
This combined fit yields the best-fit values $\check{E}_{0,CT}(\text{$r$=0})=16.00\pm0.02$~MeV, $\check{E}_{0,CT}(\text{$r$=4.5~m})=17.85\pm0.02$~MeV, $k_B=0.064\pm0.003$~mm/MeV, and $f=0.024\pm0.001$, for the spherical setup.  Again, the RMS is 0.14\%.  
Compared with the approach of independently fitting $\check{E}_{0,CT}$, the results are consistent and the errors are about twice as large for the same statistics.  However, the RMS of the data about the fit is equivalent, and compared with the fit of either individual data set in Eq.~(\ref{eq:NL2}), the estimated uncertainties are greatly reduced.  
This combined fit approach is also illustrated for the fit of the optical response, and is described in Section~\ref{sec:fitNonuniform}.

\subsection{Optical response - nonuniformity}
\label{sec:nonuniformity}

The optical response of a detector has been divided into two parts: phototransmission and photocoverage.
With the model presented here, a spatial map of the detector energy response is determined using transmission and coverage maps, and two fit parameters; namely, the energy scale $\check{E}_0$ and the effective attenuation length $L$.

\subsubsection{Phototransmission and photocoverage}

\begin{figure}[htbp] %  figure placement: here, top, bottom, or page
	\centering
		\includegraphics[width=0.49\columnwidth]{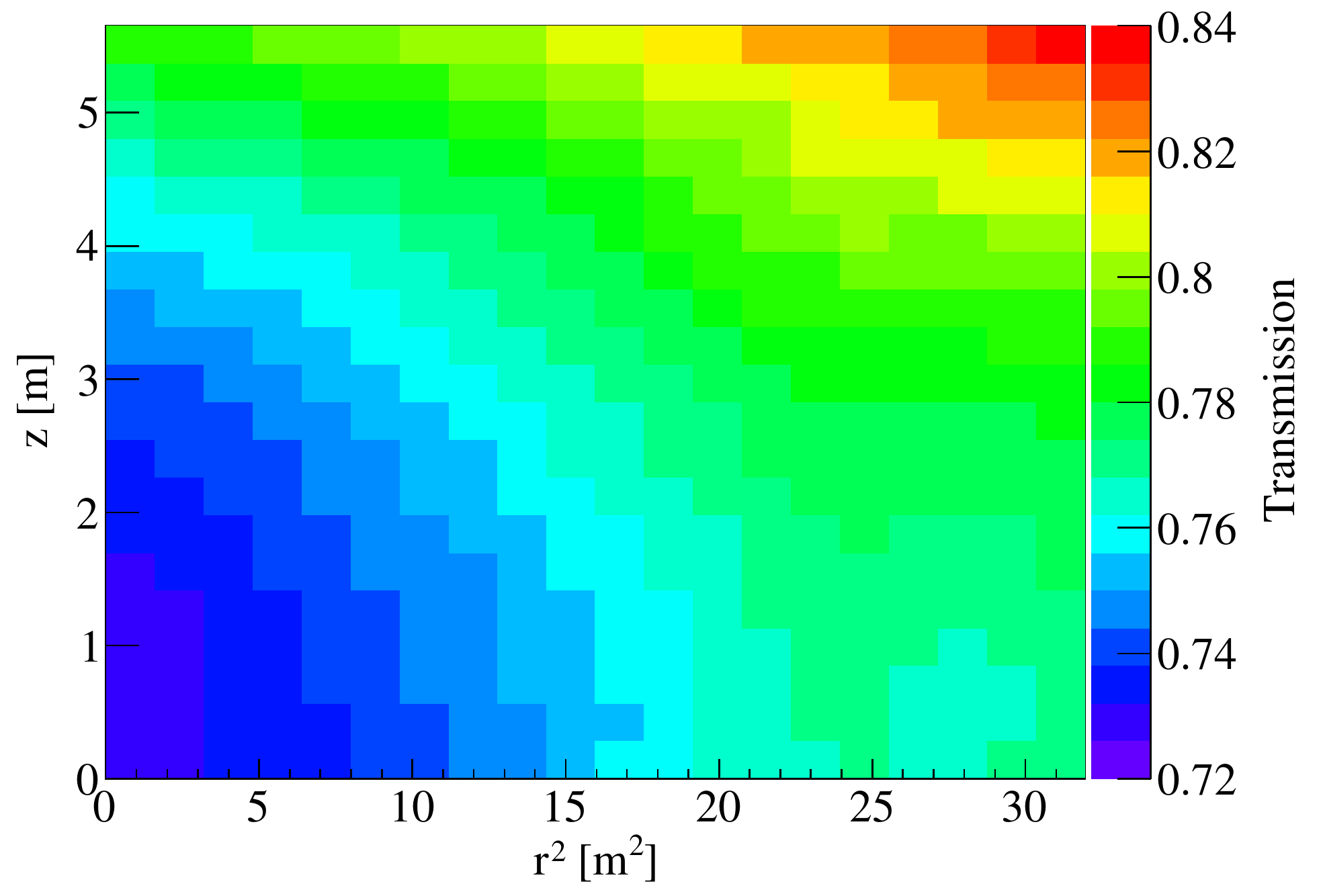}
		\includegraphics[width=0.49\columnwidth]{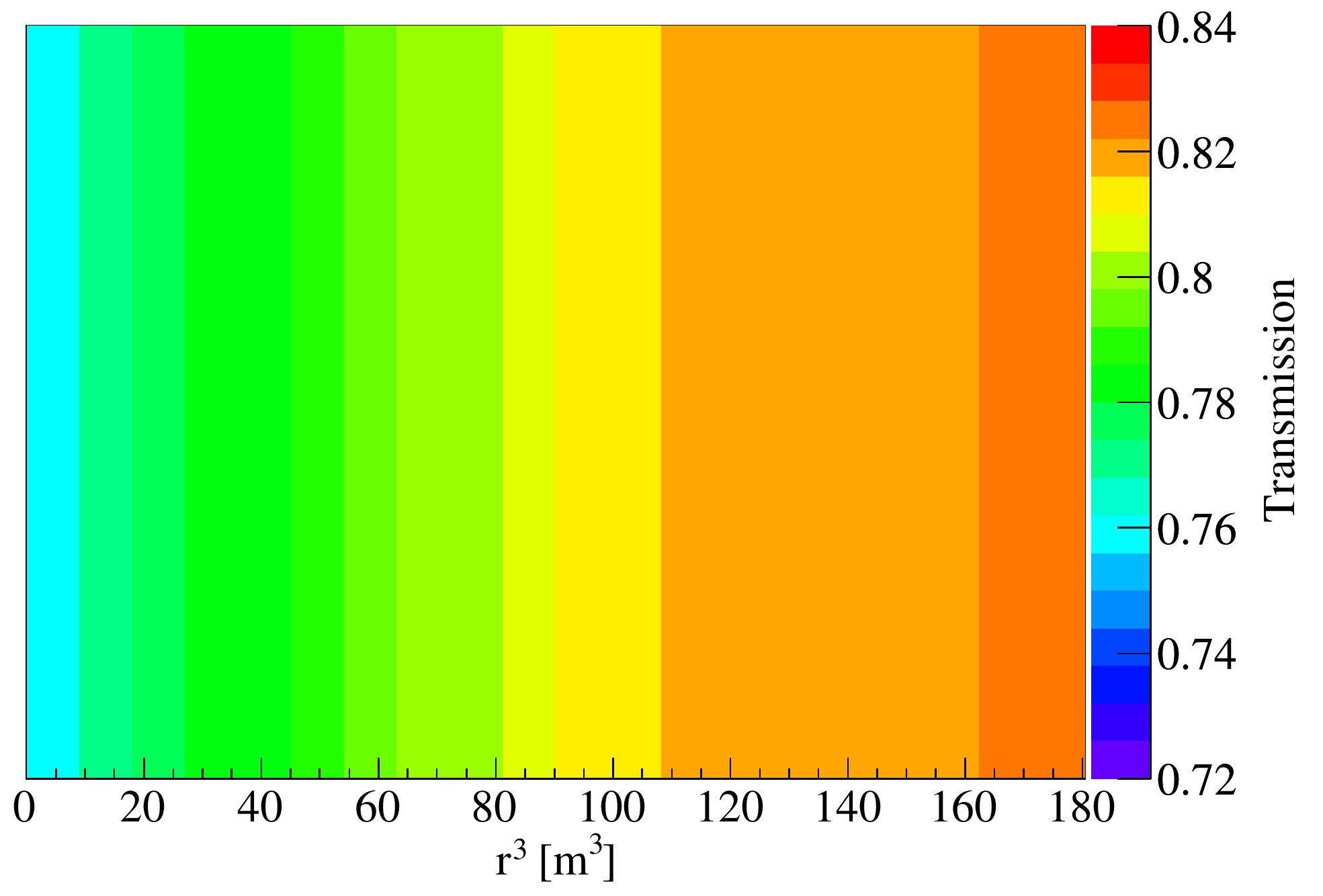}
	\caption{Phototransmission map for the cylindrical setup (left) and spherical setup (right) with $L=20.0$~m.}
	 \label{fig:trans}
\end{figure}

The transmission process is modeled using the effective attenuation length $L$ that is determined by properties of the scintillator, and a set of moment maps $m_k(\vec x)$ from Eq.~(\ref{eq:transmission}), which is unique to the detector.
Figure~\ref{fig:trans} shows the transmission maps $T(\vec x)$ of both simulated detectors assuming $L=20.0$~m.  The maps use the convention of detector center at $r = 0$ and $z = 0$.  
Here, the number of moment maps is cut at $k \le 10$, resulting in a negligible error.  
Greater transmission is observed nearer the PMTs as expected.

\begin{figure}[htbp] %  figure placement: here, top, bottom, or page
	\centering
		\includegraphics[width=0.49\columnwidth]{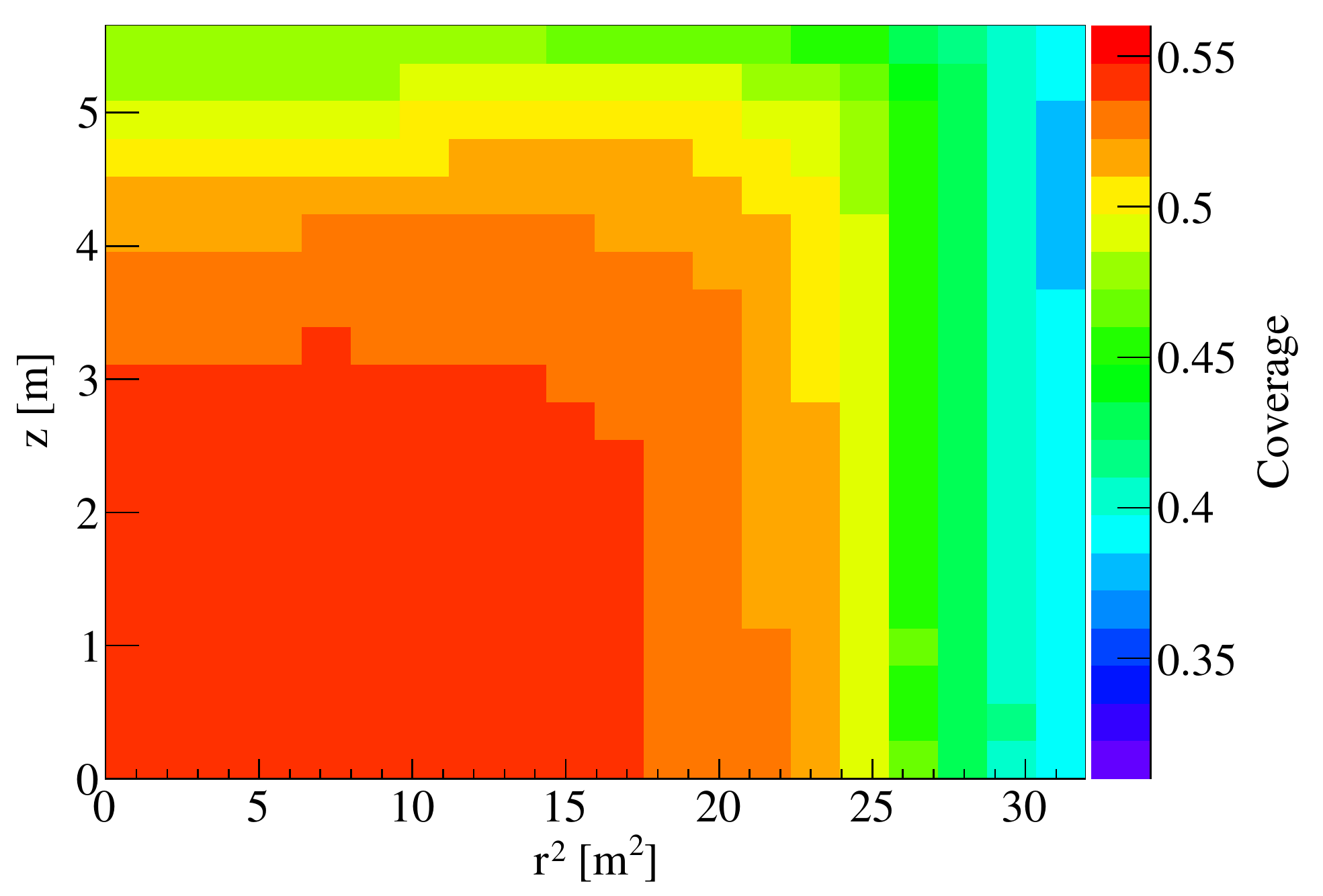}
		\includegraphics[width=0.49\columnwidth]{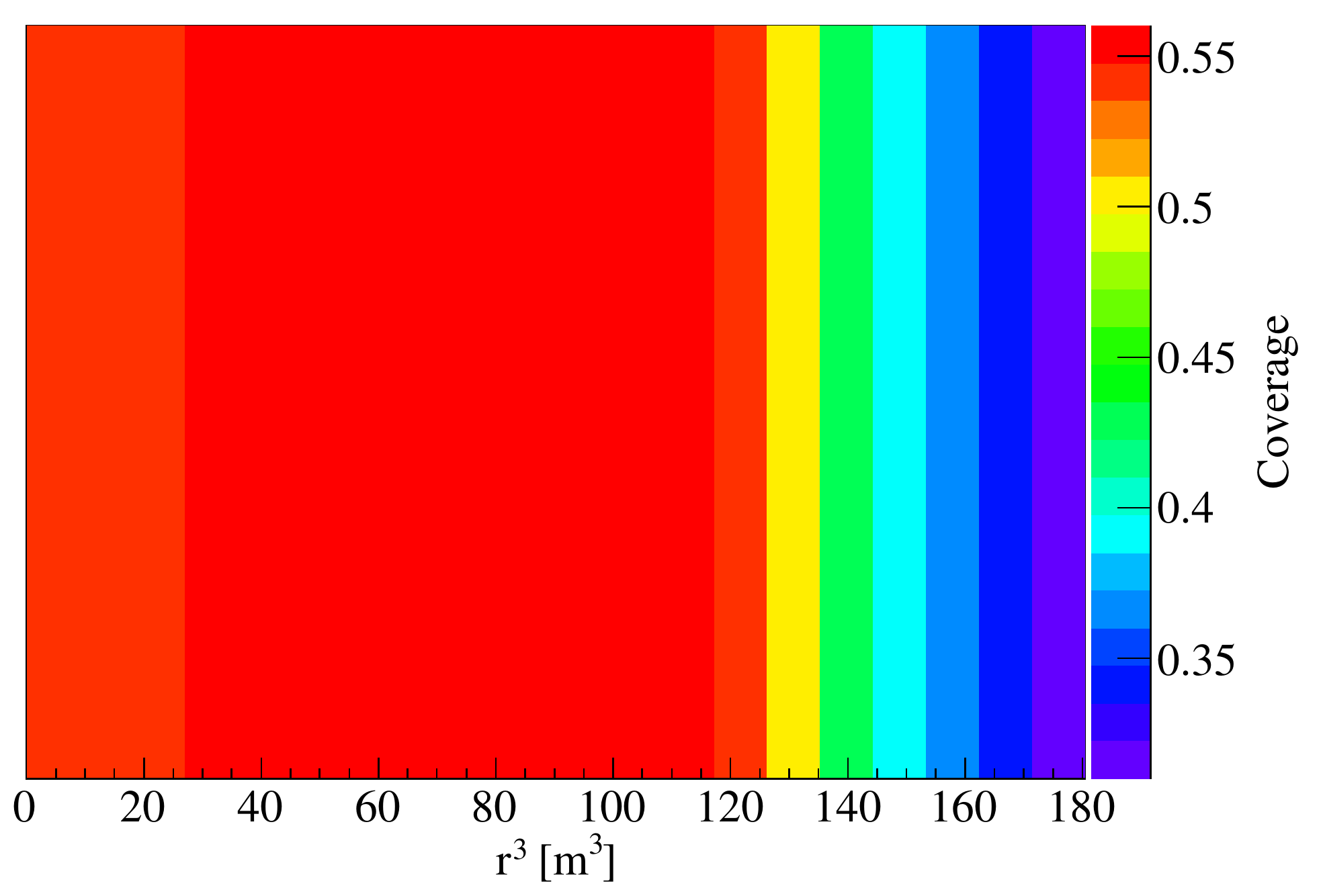}
	\caption{Photocoverage map for the cylindrical setup (left) and spherical setup (right).}
	 \label{fig:cover}
\end{figure}

Figure~\ref{fig:cover} shows the corresponding photocoverage maps $C(\vec x)$.
Counterintuitively, the coverage maps decline with increasing radius.  This is due to reflections at the acrylic/water interface, which are more significant relative to the photocoverage than to the phototransmission.  
These reflections are greatly reduced when the water is replaced with a material that has a refractive index closer to the indexes of scintillator and acrylic.  
The generation procedures for both types of maps are described in Sections~\ref{sec:transmission} and \ref{sec:coverage}, respectively.  

\begin{figure}[htbp] %  figure placement: here, top, bottom, or page
	\centering
		\includegraphics[width=0.49\columnwidth]{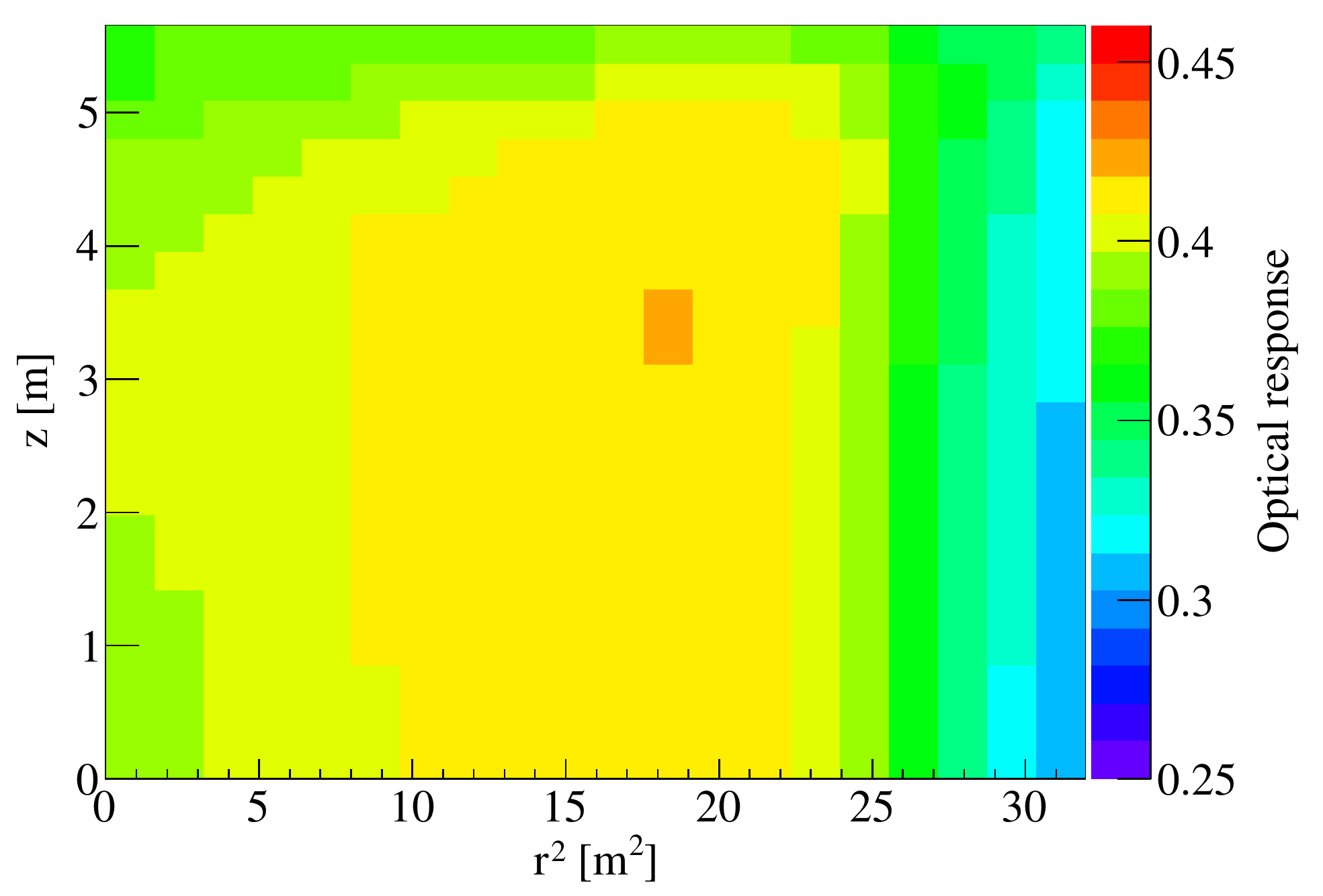}
		\includegraphics[width=0.49\columnwidth]{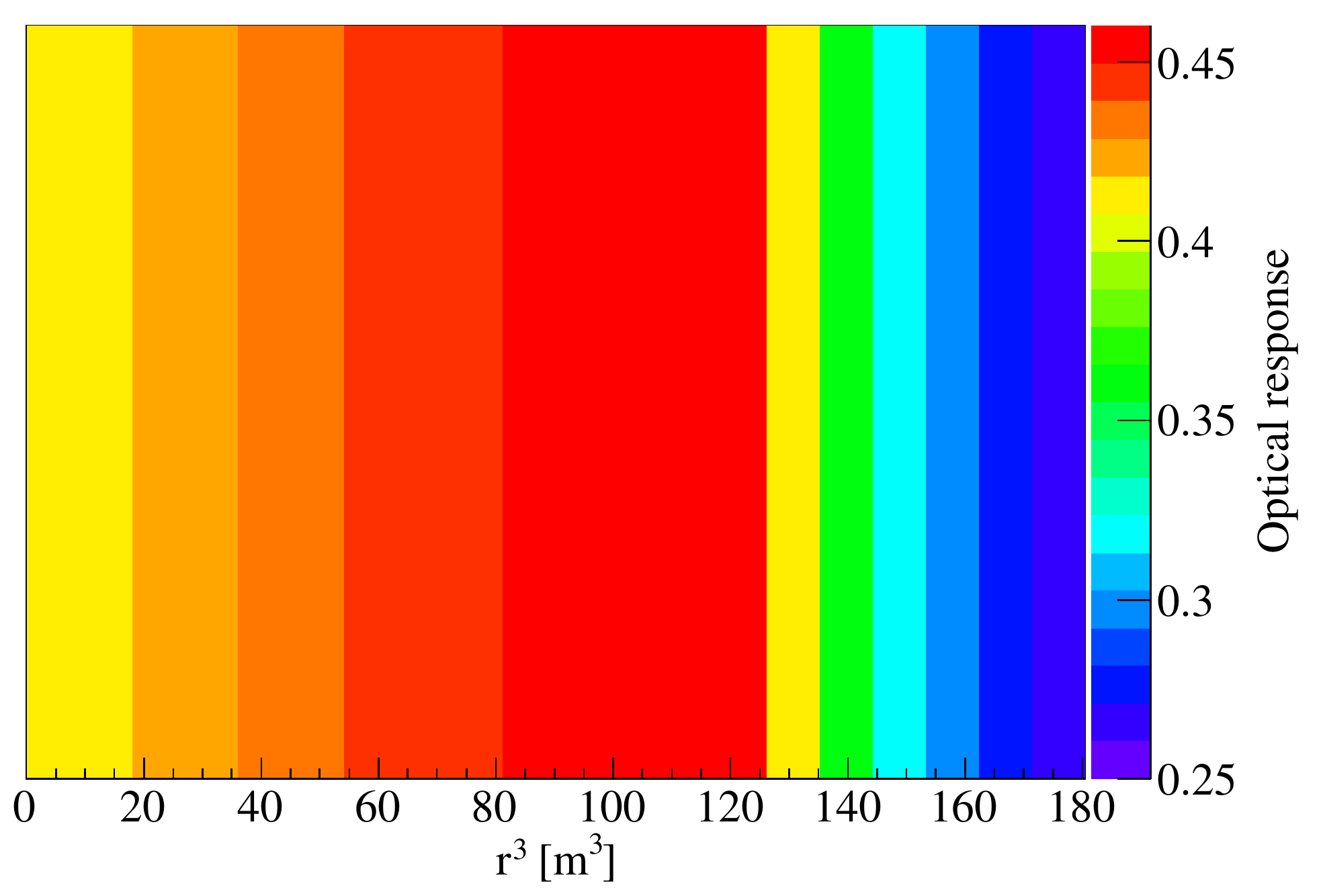}
	\caption{Optical response for the cylindrical setup (left) and spherical setup (right).}
	 \label{fig:opres}
\end{figure}

Figure~\ref{fig:opres} shows the optical response [p.e./photon] of the detectors from the full simulations using \textsc{Geant4}, considering scintillator attenuation, optical processes at the boundary of the LAB and water, and attenuation in the water buffer, which is assumed to have a 150.0~m attenuation length, averaged over photon wavelength.
The response of the two detectors is qualitatively similar in that it peaks at median radii and exhibits a minimum at largest radii.  

\subsubsection{Fitting the energy scale $\check{E}_0$ and attenuation length $L$}
\label{sec:fitNonuniform}

The energy scale $\check{E}_0$ and effective attenuation length $L$ can be simultaneously determined by fitting the observed energy $E_\mathrm{obs}$ of a mono-energetic source of energy $E_\mathrm{cal}$ distributed throughout the scintillator.  In this case, the photon yield $\hat{Y}(E_\mathrm{cal})$ is a constant, and therefore grouped with $\check{E}_0$ into a composite energy scale $E_{0,Y}(E_\mathrm{cal}) \equiv \check{E}_0\cdot \hat{Y}(E_\mathrm{cal})$, which has units of MeV/p.e.$\times$photon.  Thus, Eq.~(\ref{eq:model}) reduces to 
\begin{equation}\label{eq:NU}
	E_\mathrm{obs}(\vec x,E_\mathrm{cal})=E_{0,Y}(E_\mathrm{cal})\cdot C(\vec x)\cdot T(\vec x, L).
\end{equation}
As in the fits in Section~\ref{sec:photonYieldFit}, the photon yield is normalized to 1 at a selected energy [see Eq.~(\ref{eq:NLnorm})].  If this selected energy is $E_\mathrm{cal}$, then $E_{0,Y}=\check{E}_0$ and Eq.~(\ref{eq:NU}) becomes 
\begin{equation}\label{eq:NUnorm}
	E_\mathrm{obs}(\vec x,E_\mathrm{cal})=\check{E}_{0}\cdot C(\vec x)\cdot T(\vec x, L).
\end{equation}

\begin{figure}[htbp] %  figure placement: here, top, bottom, or page
	\centering
		\includegraphics[width=0.49\columnwidth]{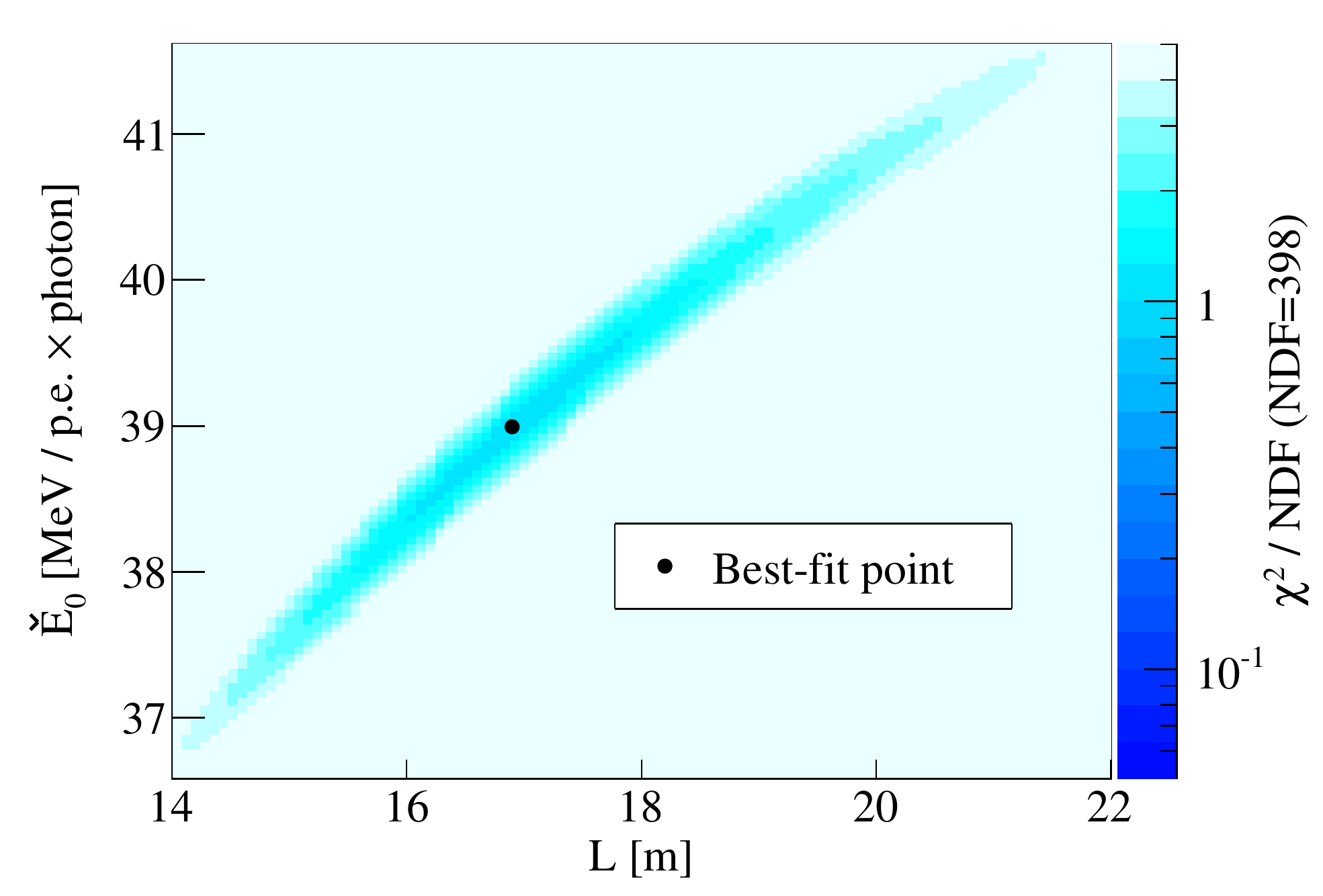}
		\includegraphics[width=0.49\columnwidth]{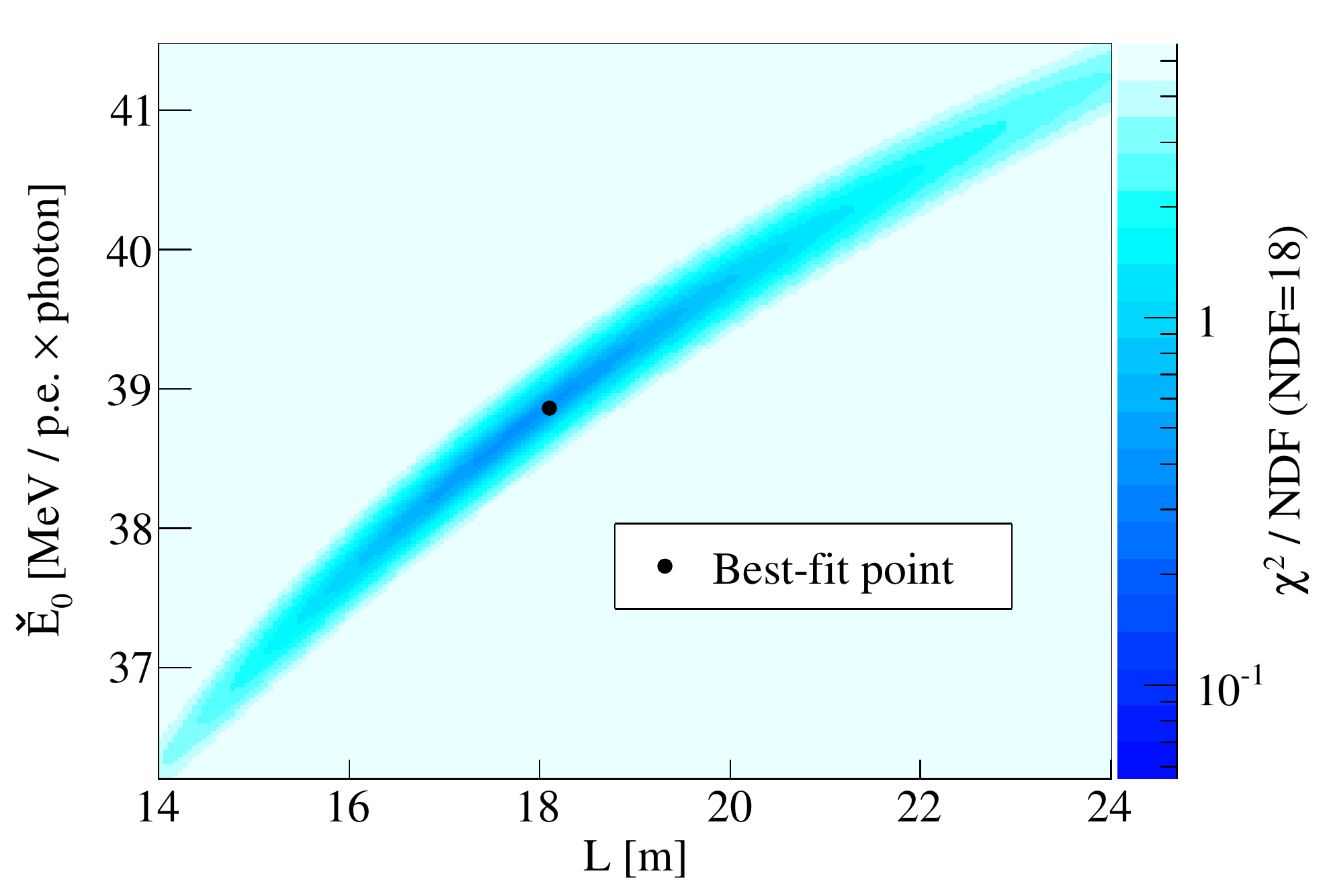}
	\caption{Sensitivity of $E_0$ and $L$ for the cylindrical setup (left) and spherical setup (right) with 16.0-MeV electrons.}
	 \label{fig:sens}
\end{figure}

Simulating 16-MeV electrons throughout the LAB, the two-parameter fit space for the two detector setups is shown in Fig.~\ref{fig:sens}.
The best-fit points are $\check{E}_0 = 38.99$~MeV/p.e.$\times$photon and $L = 16.9$~m for the cylindrical setup and $\check{E}_0 = 38.86$~MeV/p.e.$\times$photon and $L = 18.1$~m for the spherical setup.  
There is extreme anti-correlation between $\check{E}_0$ and $L$: the correlation coefficient is -0.99 for both setups. This is not a feature specific to the model, but rather, the physics.  

The ambiguity due to the extreme correlation between $\check{E}_0$ and $L$ may be minimized by fitting for $\check{E}_0$ individually.  Fitting for $L$ with $\check{E}_0$ fixed to 41.41 and 39.82~MeV/p.e.$\times$photon (which correspond to 500 and 520~p.e./MeV), gives $L =$ 19.6$\pm$0.2 and 20.1$\pm$0.3~m, for the cylindrical and spherical setups, respectively.  

Besides fitting for $\check{E}_0$ individually, the correlations can also be handled by performing a combined fit of multiple data sets of differing energy, so as to obtain multiple distinct $\check{E}_0$ and a single $L$.
In the case of a 0.5-MeV and 16.0-MeV electron source, the combined fit function would be
\begin{equation}
	\begin{aligned}
		E_\mathrm{obs}(\vec x, \text{16.0 MeV})&=E_{0,Y}(\text{16.0 MeV})\cdot C(\vec x)\cdot T(\vec x, L),
		\\	E_\mathrm{obs}(\vec x, \text{0.5 MeV})&=E_{0,Y}(\text{0.5 MeV})\cdot C(\vec x)\cdot T(\vec x, L),
	\end{aligned}
\end{equation}
where $L$ is a parameter in common.  
This combined fit yields the best-fit points $\check{E}_0 \equiv E_{0,Y}(\text{16.0~MeV})=38.99\pm0.08$~MeV/p.e.$\times$photon, $E_{0,Y}(\text{0.5~MeV})=40.12\pm0.10$~MeV/p.e.$\times$photon, and $L=16.9\pm0.1$~m for the cylindrical setup, and $E_{0,Y}(\text{16.0~MeV})=38.86\pm0.28$~MeV/p.e.$\times$photon, $E_{0,Y}(\text{0.5~MeV})=40.18\pm0.39$~MeV/p.e.$\times$photon, and $L=17.9\pm0.7$~m, for the spherical setup.
The central values agree with those from fits of either individual data set, however, the estimated uncertainties are greatly reduced.  

\subsubsection{Residual nonuniformity}

\begin{figure}[htbp] %  figure placement: here, top, bottom, or page
	\centering
		\includegraphics[width=0.49\columnwidth]{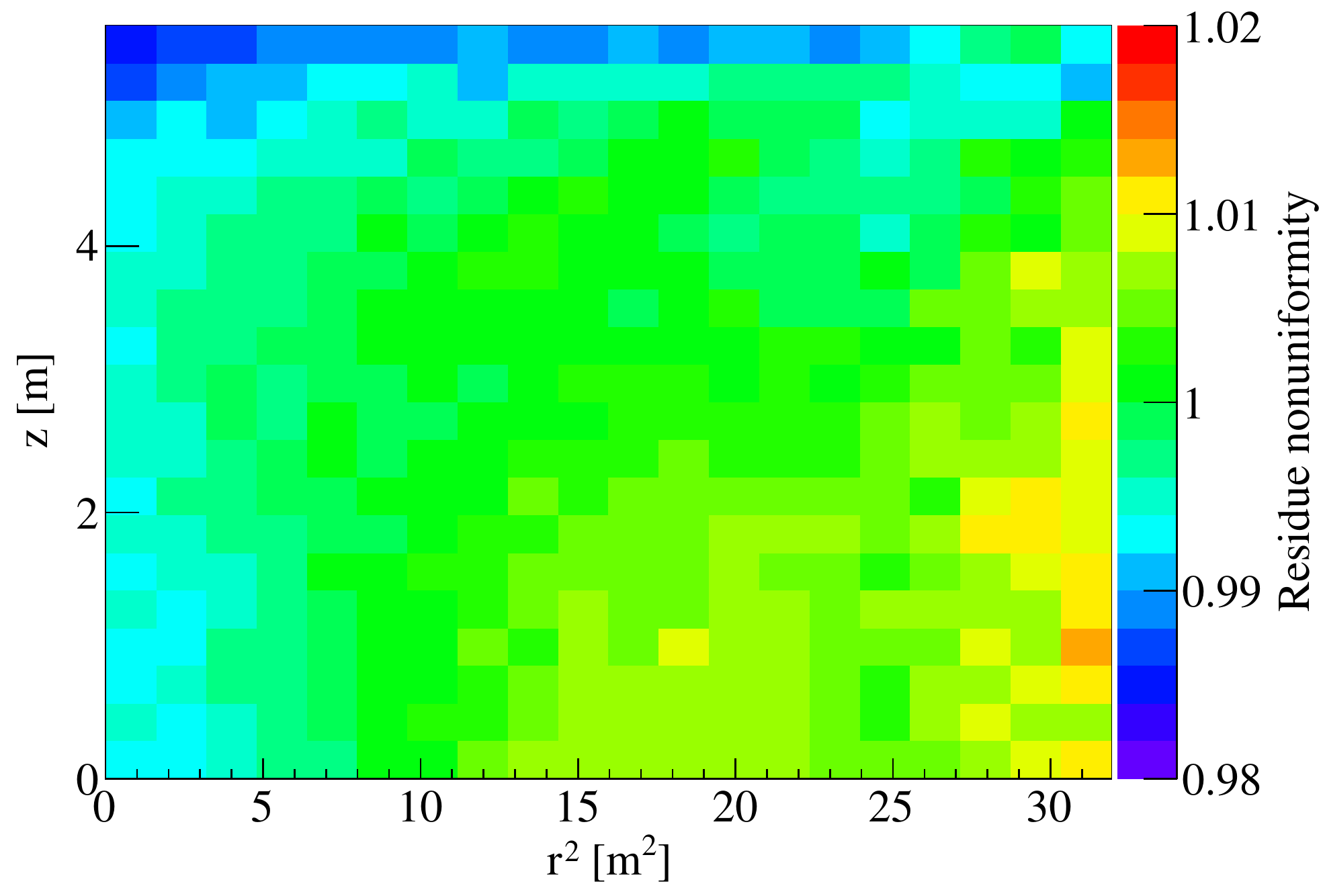}
		\includegraphics[width=0.49\columnwidth]{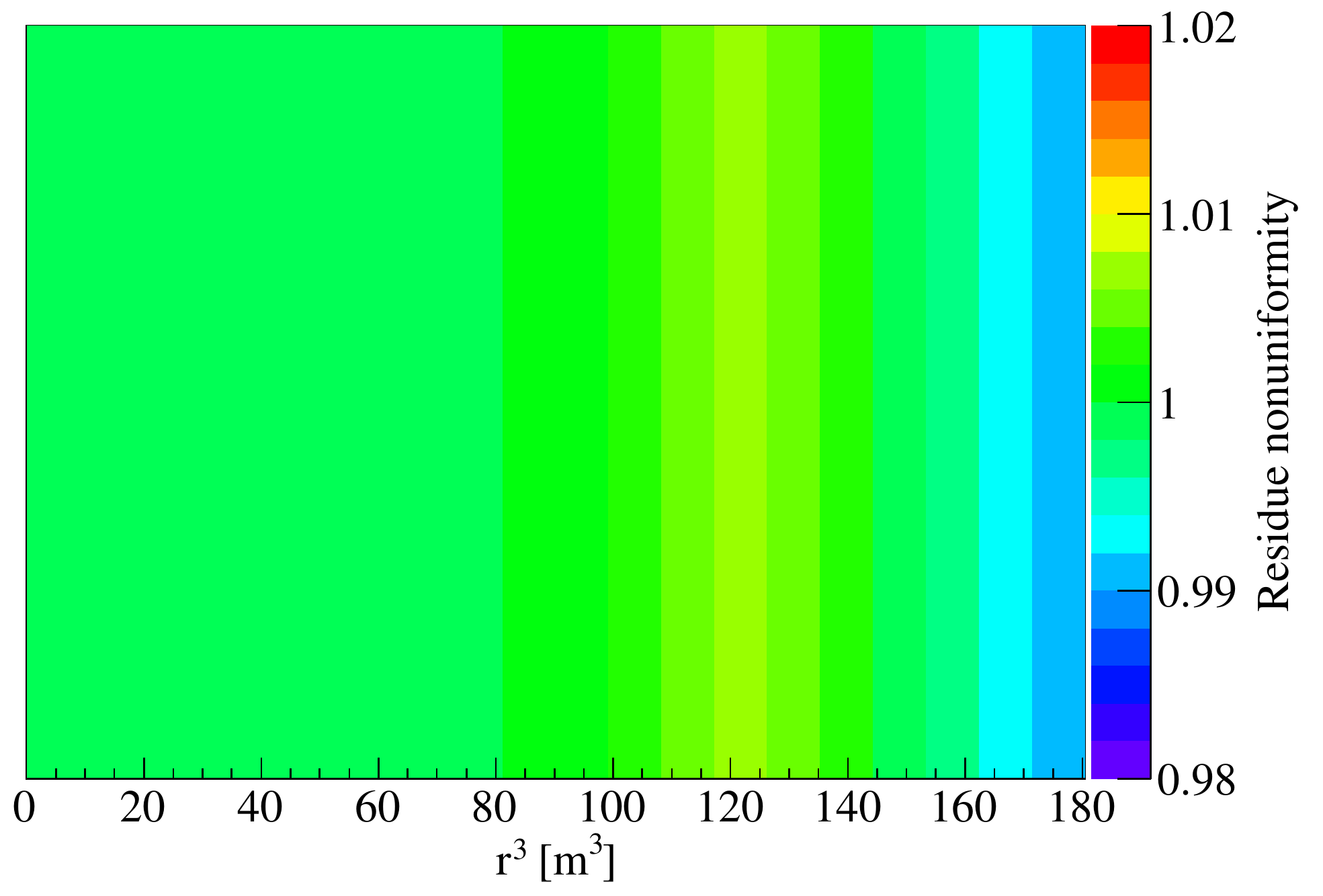}
	\caption{Residual nonuniformity for the cylindrical setup (left) and spherical setup (right).}
	 \label{fig:correct}
\end{figure}

To evaluate the fit results, a residual nonuniformity map $O^\prime(\vec x)$ is obtained:
\begin{equation}
O^\prime(\vec x)=E_\mathrm{obs}(\vec x)/[\check{E}_0\cdot C(\vec x)\cdot T(\vec x, L)],
\end{equation}
where $\check{E}_0$ and $L$ represent the best-fit values for 16-MeV electrons.  
Figure~\ref{fig:correct} shows $O^\prime(\vec x)$ for both detectors.  
The residual nonuniformity in Fig.~\ref{fig:correct} is quantified as the RMS of the values at each position bin, and is 0.37\% for the cylindrical setup and 0.26\% for the spherical setup.  

\subsection{Fitting the full model}
\label{sec:fitFullModel}

Fitting each parameter individually will most likely yield the smallest uncertainties for each parameter and better probe the validity of each model component, however, it may not provide the smallest uncertainty for the full model, it may involve complicated error propagation between the fits, and it is unlikely to minimize biases.  Therefore, fitting all parameters simultaneously and individually may both be useful, depending on the model sophistication and the calibration sources available.  
In Section~\ref{sec:photonYieldFit}, the nonlinear photon yield is fit using two distinct positions, and in Section~\ref{sec:fitNonuniform}, the nonuniform optical response is fit using two distinct energies.  
These dual-position and dual-energy fits provide a handle on the correlations among parameters and better constrain their uncertainties (compared with single-energy or single-position fits).  
This approach may be generalized to using a number of sources of differing energies and position distributions to fit the full model of Eq.~(\ref{eq:model}), simultaneously determining the energy scale $\check{E}_0$, effective attenuation length $L$, relative Cherenkov photon efficiency $f$, and Birks' constant $k_B$.  
Normalizing the photon yields to 1 at a single calibration energy, the full model of Eq.~(\ref{eq:model}) is expressed as 
\begin{equation}\label{eq:modelNorm}
	E_\mathrm{obs}(\vec x,E_\mathrm{init},E_\mathrm{dep})=\check{E}_0\cdot R \cdot C(\vec x)T(\vec x, L)\cdot \left[\frac{\hat{Y}_s(k_B)+f\hat{Y}_C}{1+f}\right](E_\mathrm{init},E_\mathrm{dep}).
\end{equation}

To demonstrate a full model fit with Eq.~(\ref{eq:modelNorm}), we use all four samples simulated for the spherical detector in the previous sections: the uniformly distributed 0.5-MeV and 16-MeV samples, and the energy scan samples at $r =$ 0 and $r =$ 4.5~m.  Simultaneously fitting all four samples yields $\check{E}_0 =$ 39.98 $\pm$ 0.02~MeV/p.e.$\times$photon, $k_B =$ 0.064 $\pm$ 0.003~mm/MeV, $f =$ 0.024 $\pm$ 0.001, and $L =$ 18.0 $\pm$ 0.1~m.  
The RMS of the residuals for the nonlinearity curve and nonuniformity map is 0.14\% and 0.40\%, respectively.  

If $\check{E}_0$ is determined first and then fixed to 39.82~MeV/p.e.$\times$photon (or 520.0~p.e./MeV), the fitting results are $k_B =$ 0.063 $\pm$ 0.001~mm/MeV, $f =$ 0.024 $\pm$ 0.001, and $L =$ 18.0 $\pm$ 0.1~m.  The residual RMS is 0.15\% and 0.44\%, respectively.

\section{Discussion on systematic uncertainty}
\label{sec:systematics}

The presented model describes the primary components of detector response with a minimal set of physically-motivated parameters that are fit to data.  By working with physically-motivated model components, potential deficiencies of the detector response description may be more directly revealed, for example, from uncorrected features of calibration sources or from isolated regions of inadequately modeled geometry or material properties. 
Along the same lines, these model parameters make it easy to study the systematics of the detector response, and their correlations.  
Notably, the parameters can naturally capture changes of liquid scintillator properties over time.

\subsection{Efficient study of systematic uncertainties}

Event selection efficiencies and associated systematics can be investigated efficiently by performing analytical Monte Carlo calculations using the presented model instead of full simulations, such as those using \textsc{Geant4}.  
The model, as expressed by Eq.~(\ref{eq:model}), can make use of a particle's initial and deposited energies, $E_\mathrm{init}$ and $E_\mathrm{dep}$, and event position $\vec x$, which need be generated only once from an independent calculation or simulation.  
With this single event sample in hand, the fit parameters of the model, or other quantities, such as position resolution and bias, and fiducial volume selections, can be directly studied, along with their correlations.   
Though energy resolution is not discussed here (see, for example, Ref.~\cite{Logan:2016}), it can be obtained from the observed energy in data and then applied as a smearing at the end of the calculation.  
Thus, the model has full capability for studying systematic effects on energy spectra without a need to generate various event samples.  

\subsection{Systematic uncertainties from the time-variation of scintillator}
\label{sec:sysTime}

Scintillator experiments generally calibrate their energy scale versus time due to expected changes in the scintillator.  
Two examples of a change of energy scale over time are -1.5\% over 4 years in KamLAND~\cite{Hiroko:2011} and -0.5\% to -2.0\% per year in Daya Bay~\cite{An:2016ses}.
Not simply due to a decrease in photon yield, part of the change in these energy scales is due to the decrease of the attenuation length of the scintillator.  
This implies a change in the nonuniformity of the detector response, and could produce a bias of the energy scale if uncorrected.
Indeed, scintillator experiments often treat this change in nonuniformity as a systematic uncertainty.  
Applying Eq.~(\ref{eq:transFraction}) to data in Ref.~\cite{An:2016ses}, Daya Bay's effective attenuation length $L$ decreases by about 3\% per year.
In some cases, particularly those with doped or loaded scintillator, this rate is an order of magnitude greater~\cite{Apollonio:2002gd, Kim:2016tyv}.  

\begin{figure}[htbp] %  figure placement: here, top, bottom, or page
	\centering
		\includegraphics[width=0.49\columnwidth]{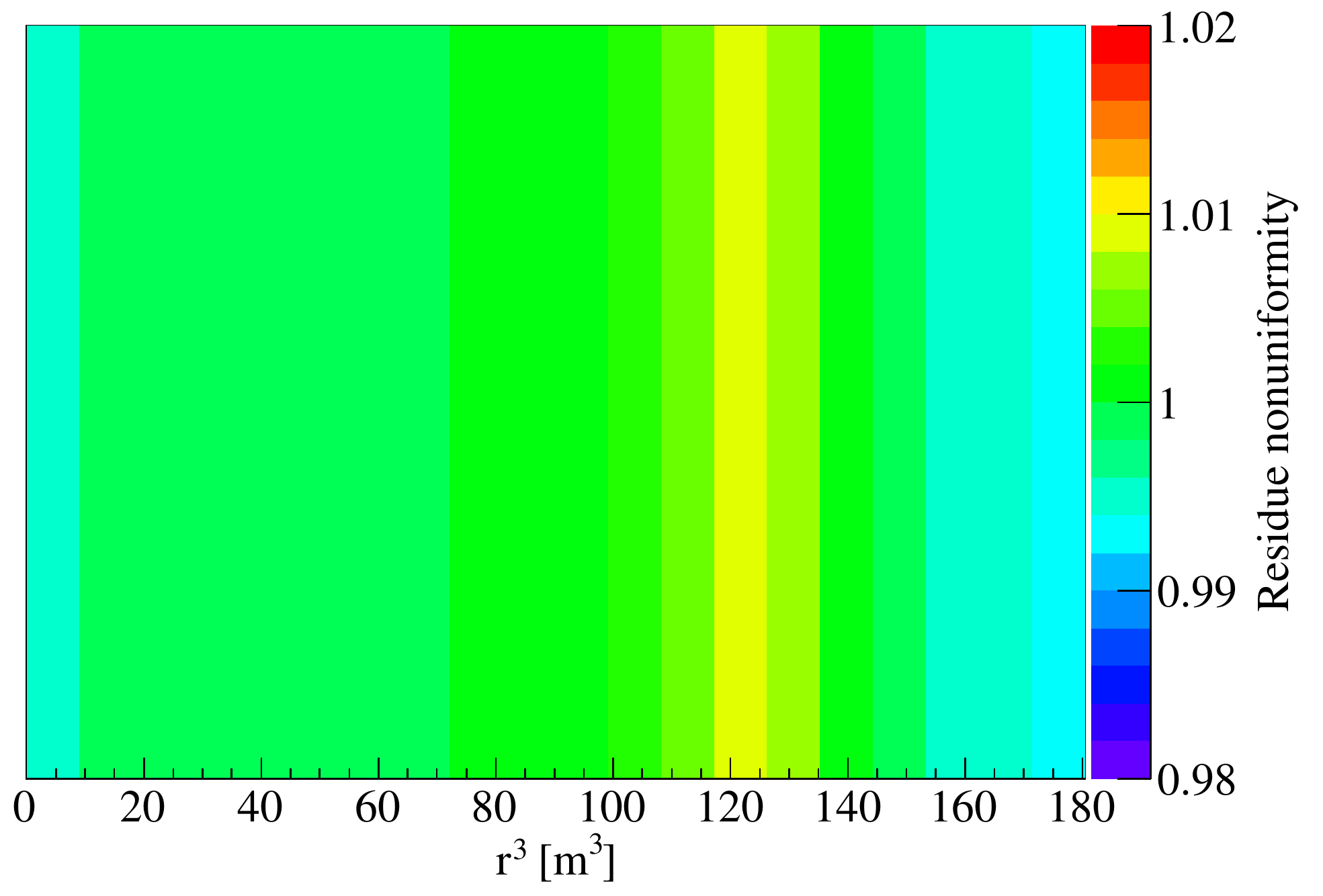}
		\includegraphics[width=0.49\columnwidth]{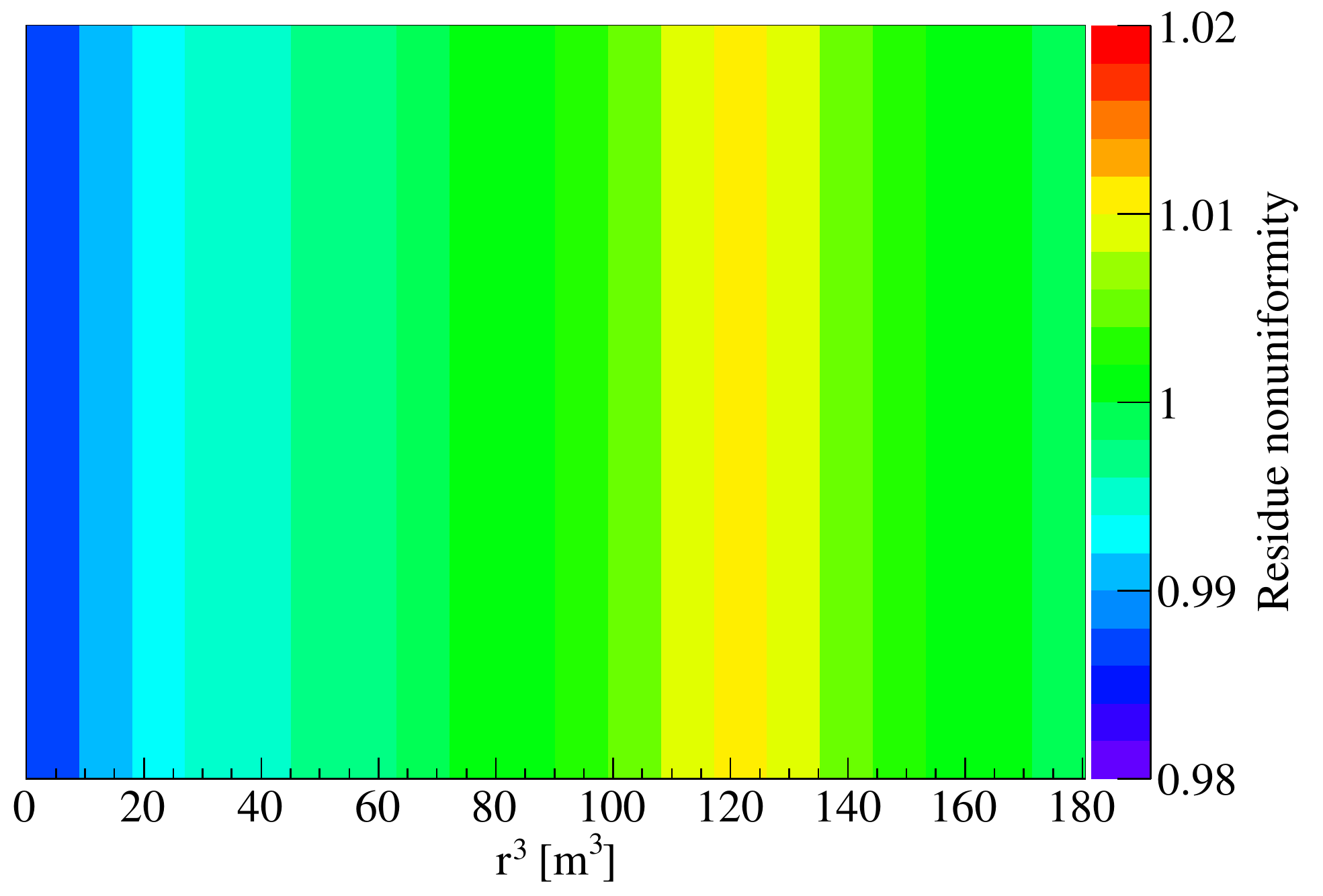}
	\caption{Residual nonuniformity maps for the spherical setup after five years, fitting the increase in scintillator attenuation length over time (left), and fitting only a single attenuation length (right).}
	 \label{fig:t5}
\end{figure}

To illustrate the impact of neglecting the time variation of $L$, we consider the spherical detector described in Section~\ref{sec:validate}, with $\check{E}_0 =$~39.82~MeV/p.e.$\times$photon and $L =$~20~m at initial time $t_0$, and $\check{E}_0 =$~35.99~MeV/p.e.$\times$photon and $L =$~17~m at time $t_5$, five years after $t_0$.  Fitting these two detector states with the presented model produces the residual nonuniformity maps shown in Fig.~\ref{fig:correct} ($t_0$) and the left panel of Fig.~\ref{fig:t5} ($t_5$).  The RMS of the two maps is 0.26\% and 0.28\%, respectively.  
If $L$ is not fit over time, but fit at only $t_0$, then the residual nonuniformity at $t_5$ has an RMS of 0.45\%, as illustrated in the right panel of Fig.~\ref{fig:t5}.  
A more significant increase occurs for the cylindrical detector: 0.4\% would increase to 0.8\%.  
Thus, the modeling of the time variation of both $\check{E}_0$ and $L$ as demonstrated in Section~\ref{sec:nonuniformity}, would generally avoid an additional systematic uncertainty or bias of energy estimates.  

For multi-volume scintillator detectors such as the gadolinium-loaded Daya Bay detectors or the enriched xenon gas-loaded KamLAND-Zen detector, distinct attenuation lengths for the loaded volume and the pure scintillator volume can be fitted as described by Eq.~(\ref{eq:2vtrans}), accounting for the distinct time variations of the different volumes.

\section{Conclusion}
\label{sec:conclusion}

The generic energy response model of liquid scintillator detectors presented in this study can be computed quickly and can provide immediate energy estimations of better than 0.5\% accuracy over the typical lifetime of a kiloton-scale experiment. 
The model allows the individual or simultaneous fit of the energy scale $\check{E}_0$, effective attenuation length $L$, relative Cherenkov efficiency $f$, and Birk's constant $k_B$, to experimental data, and can accommodate multi-volume detectors. 
The minimal set of physically-motivated parameters captures the essential characteristics of scintillator response and can naturally account for changes in scintillator over time, helping to avoid associated biases or systematic uncertainties.
Furthermore, the model provides an efficient framework for quantifying systematic uncertainties and correlations.   

\section{Acknowledgments}
This work is supported in part by, the National Natural Science Foundation of China (Nos. 11475093 and 11620101004), the program of China Scholarships Council No. 201606210394, the Key Laboratory of Particle \& Radiation Imaging (Tsinghua University), and the CAS Center for Excellence in Particle Physics (CCEPP). 

\section*{References}




%\bibliographystyle{elsarticle-num}
%\bibliography{detectorResponse}

\end{document}